\documentclass[11pt,preprintnumbers,aps,amssymb,nofootinbib,amsmath,superscriptaddress]
{revtex4}
\usepackage{epsfig,epsf,comment}
\usepackage{bm} 
\usepackage{slashed}
\usepackage{color} 
\newcommand{\beq}{\begin{equation}}
\newcommand{\beql}[1]{\begin{equation}\label{#1}}
\newcommand{\eeq}{\end{equation}}
\def\bal#1\gal{\begin{align}#1\end{align}}
\newcommand{\ball}[1]{\bal\label{#1}}
\newcommand{\eq}[1]{(\ref{#1})}
\newcommand{\fig}[1]{Fig.~\ref{#1}}
\renewcommand{\sec}[1]{Sec.~\ref{#1}}
\newcounter{topiccounter}
\renewcommand{\b}[1]{{\bm #1}} 
\newcommand{\unit}[1]{\hat {{\bm #1}}} 

\newcommand{\e}{\varepsilon}
\newcommand{\aver}[1]{\left\langle #1 \right\rangle}

\begin{document}

\title{Electromagnetic field of ultrarelativistic charge in topologically random nuclear matter }
\author{Kirill Tuchin}

\affiliation{Department of Physics and Astronomy, Iowa State University, Ames, Iowa, 50011, USA}

\date{\today}


\begin{abstract}
 
Electromagnetic field of a fast electric charge in nuclear matter with spatially uniform but random topological charge density is derived. A useful approximation is developed for the relativistic heavy-ion collisions.

\end{abstract}

\maketitle

\section{Introduction}\label{sec:i}

Electromagnetic field of a fast charge moving through the topologically nontrivial matter is of interest in  many physics fields. A solution to this problem was found when the topological charge density is nearly stationary and homogenous and the electromagnetic response is dominated by low frequencies  \cite{Tuchin:2014iua,Li:2016tel}, which is believed to be relevant to the phenomenology of the relativistic heavy-ion collisions as long as the considered time intervals as shorter than the sphaleron transition time $\tau_c$ \cite{Arnold:1997gh}. It was shown in \cite{Tuchin:2014iua,Li:2016tel} that the finite topological charge can have a profound effect on the electromagnetic field. The main goal of this paper is to compute  the electromagnetic field  in the opposite limit viz.\  at time intervals much longer than $\tau_c$ by treating the topological charge density as a spatially uniform \cite{Zhitnitsky:2014ria,Kharzeev:2007tn}  stochastic process. This approach is developed in my recent paper \cite{Tuchin:2019gkg} where it is applied to investigate the late-time behavior of the chiral instability.

The presentation is structured as follows. In \sec{sec:a} the Maxwell-Chern-Simons (MCS) equations are expanded in the helicity basis which is particularly convenient for discussing the topological effects. In particular, the effect of the time-dependence of the topological charge can be computed by solving an ordinary differential equation \eq{a18}. The kinematics of the relativistic heavy-ion collisions allows for a few very accurate approximations of MCS equations, which is reviewed in \sec{sec:k}. The electromagnetic field at early times $t\ll \tau_c$ is discussed in \sec{sec:b} were the results of \cite{Tuchin:2014iua} are reproduced in the helicity basis. The solution to MCS equations at later times is derived in \sec{sec:b2} by ensemble averaging equation \eq{a18}.  The main result is displayed in \eq{D1} and \fig{fig:2}. Summary is presented in \sec{sec:s}.

\section{Electromagnetic field in the helicity basis }\label{sec:a}

Electrodynamics in the topological matter is described using the Maxwell-Chern-Simons theory  which adds to the Maxwell Lagrangian a term that couples $F\tilde F$ directly to the topological charge density\cite{Wilczek:1987mv,Carroll:1989vb,Sikivie:1984yz}. The field equations for a point charge $q$ moving in the positive $z$ direction with constant velocity $v$ read
\begin{subequations}\label{A1}
\bal
&\b \nabla \times \b B = \partial_t \b D +\sigma_\chi \b B+ \b j \,,\label{a1}\\
&\b \nabla\cdot \b D= \rho\,,\label{a2}\\
&\b \nabla \times \b E =-\partial_t \b B\,,\label{a3}\\
&\b \nabla\cdot \b B=0\,,\label{a4}
\gal 
\end{subequations}
where the external current is $ j^\mu = (\rho,\b j)=  q(1,v\unit z) \delta(z-vt)\delta (\b b)$, $z$ and $\b b$ are the longitudinal and transverse components of the position vector. The displacement is given by 
\ball{a5} 
\b D(\b r, t) = \int_{0}^\infty\varepsilon (t')\b E(\b r, t-t')dt'\,.
\gal
The spectral representation of permittivity is assumed to be $\varepsilon_\omega = 1+i\sigma/\omega$, where $\sigma$ is electrical conductivity (taken to be a constant). The chiral conductivity $\sigma_\chi(t)$ is proportional to the time-derivative of the topological charge density of matter. It is modeled as stationary $\sigma_\chi(0)$ at $t\ll \tau_c$, whereas at $t\gg \tau_c$  as a  stochastic process with vanishing expectation value $\aver{\sigma_\chi(t)}=0$ and the dispersion $\Sigma_\chi=\sqrt{\aver{\sigma_\chi^2}}=\sigma_\chi(0)$.

Due to the presence of the anomalous current $\sigma_\chi \b B$ in \eq{a1}, it is natural to seek a solution of Eqs.~\eq{A1} as a superposition of the helicity states, which are the eigenstates of the curl operator in the Cartesian coordinates, 
\begin{subequations}\label{A6}
\bal
  \b B(\b r,t)&= \sum_\lambda\int \frac{d^3k}{(2\pi)^3}e^{i\b k\cdot \b r}\b\epsilon_{\lambda \b k}\Phi_{\lambda\b k}(t)\,,\label{a6}\\
 \b E(\b r,t)&= \sum_\lambda\int \frac{d^3k}{(2\pi)^3}e^{i\b k\cdot \b r} \b\epsilon_{\lambda \b k}\Psi_{\lambda\b k}(t)+ \int \frac{d^3k}{(2\pi)^3}e^{i\b k\cdot \b r}\unit k \Psi'_{\lambda\b k}(t)\,, \label{a7}
 \gal
 \end{subequations}
 where $\lambda=\pm 1$ is helicity and  $\b\epsilon_{\lambda \b k}$ are the circular polarization vectors satisfying the orthonormality conditions  $\b\epsilon_{\lambda \b k}\cdot\b\epsilon_{\mu \b k}^*=\delta_{\lambda\mu}$, $\b\epsilon_{\lambda \b k}\cdot \b k=0$ and the identity
\ball{a8}
 i\unit k \times\b \epsilon_{\lambda \b k }= \lambda\b \epsilon_{\lambda \b k }\,.
\gal
We also expand the external current in this basis
 \ball{a10}
 \b j (\b r,t)=  \sum_\lambda\int \frac{d^3k}{(2\pi)^3}e^{i\b k\cdot \b r} \b\epsilon_{\lambda \b k}J_{\lambda\b k}(t)+ \int \frac{d^3k}{(2\pi)^3}e^{i\b k\cdot \b r}\unit k J'_{\lambda\b k}(t)\,,
 \gal
 where 
 \ball{a11}
 J_{\lambda\b k}= qv \b \epsilon_{\lambda \b k}^*\cdot \unit z e^{-ik_zvt}\,,\qquad J'_{\lambda\b k}= qv \frac{k_z}{k} e^{-ik_zvt}\,,
 \gal
which can be verified using the equations derived in Appendix.

Plugging  \eq{a7} and \eq{a5} into \eq{a2}  yields the equation:
\ball{a12}
 \int_0^\infty dt' \e(t') ik \Psi'_{\lambda k}(t-t')= q e^{-ik_z vt}\,.
 \gal
Fourier-transforming \eq{a12} with respect to time one  obtains
 \ball{a13}
 ik\Psi'_{\lambda \b k}(t)= \int_{-\infty}^\infty d\omega e^{-i\omega t}\frac{q}{\e_\omega}\delta(\omega-k_z v)= \frac{q}{\e_{k_zv} }e^{-ik_zvt}\,.
 \gal
Substituting \eq{a13} into \eq{a7}, gives the noise-free longitudinal component of the electric field.  
 
Turning to the transverse components of the field,  write \eq{a1} and \eq{a3} in momentum space
\begin{subequations}
 \bal
& i\b k\times \b \epsilon_{\lambda \b k} \Phi_{\lambda \b k}= \b\epsilon_{\lambda \b k}\dot \Psi_{\lambda \b k}+\unit k \dot \Psi'_{\lambda \b k} + \sigma  \b \epsilon_{\lambda \b k}\Psi_{\lambda \b k}+\sigma \unit k \Psi'_{\lambda \b k}+\sigma_\chi  \b \epsilon_{\lambda \b k}\Phi_{\lambda \b k}+ \b \epsilon_{\lambda \b k} J_{\lambda \b k}+ \unit k J'_{\lambda \b k}\,,\label{a14}\\
&i\b k \times \b \epsilon_{\lambda \b k} \Psi_{\lambda \b k} = -\b \epsilon_{\lambda \b k} \dot \Phi_{\lambda \b k} \label{a15}\,.
 \gal
 \end{subequations}
Using \eq{a8} in \eq{a15} yields
\ball{a17}
k \Psi_{\lambda \b k}= -\lambda\dot \Phi_{\lambda \b k}\,, 
 \gal
 while \eq{a14} reduces to the following two equations 
 \begin{subequations}
 \bal
& \lambda \ddot \Phi_{\lambda \b k}+\lambda\sigma \dot \Phi_{\lambda \b k} +(\lambda k^2-\sigma_\chi k)\Phi _{\lambda \b k}=k J_{\lambda \b k}\,,\label{a18}\\
& \dot \Psi'_{\lambda \b k}+\sigma \Psi'_{\lambda \b k} = -J'_{\lambda \b k}\,.\label{a19}
 \gal 
 \end{subequations}
 Eq.~\eq{a19} is the momentum space representation of the continuity equation $\b \nabla\cdot \b j= - \dot \rho = -\b\nabla\cdot  \dot {\b D}$, while \eq{a18} determines the magnetic field.  Solution to \eq{a18} depends on the functional form of $\sigma_\chi(t)$.

 \section{Anomaly-free solution and the heavy-ion collision kinematics}\label{sec:k}
 
Analytical solutions of \eq{a18} are derived in  \sec{sec:b} and \sec{sec:b2}  in the form of the three-dimensional Fourier integrals for constant and stochastic chiral conductivity respectively. These integrals cannot be taken exactly, but can be approximated in appropriate kinematics. This paper specifically focuses on the kinematics of the heavy-ion collisions. 

For the maximum clarity, it is instructive to discuss the heavy-ion collision kinematics using the anomaly-free solution, i.e.\ the case $\sigma_\chi=0$. The solution in this case  can be easily obtained from \eq{a18} and is well-known:
\ball{k1}
B_\phi(\b r,t)&= \int \frac{d^3k}{(2\pi)^3}e^{ik_z(z- vt)}e^{i\b k_\bot\cdot \b b}\frac{(-i\b k\cdot \unit b) qv }{k_z^2/\gamma^2+k_\bot^2-i\sigma k_z v }\,,
 \gal
where $\gamma= (1-v^2)^{-1/2}$. Integration over $k_z$ picks up one of the two poles depending on the sign of the variable $\zeta = vt-z$: 
\ball{k2}
k_{z0}^\pm =  \frac{i\sigma v\gamma^2}{2}\left( 1\pm \sqrt{1+Y}\right)\,,
\gal
where a shorthand notation is used:
\ball{k3}
Y=\frac{4k_\bot^2}{v^2\sigma^2\gamma^2}\,.
\gal
In the limit $Y\gg 1$ the poles \eq{k2} become $ k_{z0}^\pm=\pm i \gamma k_\bot$ and integration in \eq{k1} yields
\ball{k10}
B_\phi(\b r,t)&= \frac{qv\gamma}{4\pi}\frac{b}{\left(b^2+\gamma^2\zeta ^2\right)^{3/2}}\,,
\gal
which is the magnetic component of the Coulomb field of a moving charge in free space. In the opposite limit $Y\ll 1$, $k_{z0}^-= -i k_\bot^2/v\sigma$ and $k_{z0}^+=i\sigma v\gamma^2$. The corresponding magnetic field is 
 \ball{k12}
B_\phi(\b r,t)&= \frac{q\sigma b}{8\pi\zeta^2} \exp\left\{-\frac{b^2 \sigma}{4\zeta}\right\}\theta(\zeta)\,.
  \gal
The contribution at $\zeta<0$ is proportional to $\exp(-\sigma v\gamma^2|\zeta|)$. It can be neglected because in relativistic heavy-ion collisions, the valence charges move with \emph{ultrarelativistic velocities} \texttt{(i)} $\gamma \gg 1$. 

The second condition stems from the fact that the \emph{plasma is a poor electrical conductor} in the sense that $b\sigma\ll 1$. Indeed, plasma conductivity at the critical temperature is $\sigma = (34\,\text{fm})^{-1}$ \cite{Aarts:2007wj,Ding:2010ga,Amato:2013oja} and  $b=1- 10$~fm.  Since the fields depend on the transverse direction through the phase factor $e^{i\b b\cdot \b k_\bot}$, the typical transverse momentum is $k_\bot \sim 1/b$ implying the second condition \texttt{(ii)}  $k_\bot \gg \sigma$. It is seen from \eq{k2} that these two conditions imply that $k_z\gg k_\bot$, hence $k\approx k_z$. 
The numerical calculation shown in \fig{fig:2} is done in this approximation.  
  
One can develop a simple, but remarkably accurate, analytical approximation using the fact that the time-dependence of the field is given by $e^{-ik_z\zeta}$. This implies that at late times $\zeta\gg 1/\sigma\gamma$ the longitudinal component of the field spectrum is bounded by \texttt{(iii)} $k_z\ll \sigma\gamma$. In this case  $k_\bot \ll k_z\ll \gamma\sigma$ which corresponds to $b\sigma \gamma\gg 1$ or $Y\ll 1$. This condition holds well for $\gamma \sim 100$ and above. Moreover, $1/\sigma\gamma$ is smaller than the Compton wavelength of the color charges making up the plasma immediately upon the collision, thus effectively allowing us to regard $1/\sigma\gamma$ as zero (i.e.\ the collision instant). The solution in this case is given by \eq{k12}. This is the phenomenologically relevant case. 

At  not very high energies and/or small enough $b$ the opposite limit $b\sigma\gamma \ll 1$  or $Y\gg 1$ occurs. In this case there is an interval of momenta $\gamma^2\sigma \ll k_z\ll k_\bot^2/\sigma$ corresponding to the times $\sigma b^2\ll \zeta\ll 1/\gamma^2\sigma$ when the effect of the conductivity is negligible and one recovers  \eq{k10}. 

In summary, the kinematics of a typical relativistic heavy-ion collision with $\gamma>100$ is such that the field components satisfy $\sigma \ll k_\bot \ll k_z\ll \sigma\gamma$. The same conclusion holds for anomalous plasma provided that $\sigma_\chi$ and $\Sigma_\chi$ are not much larger than $\sigma$. For numerical calculation in \sec{sec:b2} only $\omega \approx k\approx k_z$ approximation is used. 
 
\section{Electromagnetic field at early times} \label{sec:b}


 At $t\ll \tau_c$, the time-evolution of the  chiral conductivity can be neglected. The time-dependence of the magnetic field follows from \eq{a11} and is given by $\Phi_{\lambda \b k}\sim e^{-ik_z vt}$. Thus, \eq{a18} and \eq{a17} yield
 \begin{subequations}
\bal
 \Phi_{\lambda \b k}&= \frac{qv \unit z \cdot\b \epsilon_{\lambda\b k}^*\lambda ke^{-ik_z vt}}{k^2-\lambda \sigma_\chi k -(k_zv)^2-i\sigma k_z v}\,,\label{a22} \\
 \Psi_{\lambda \b k}&=\frac{iqk_zv^2 \unit z \cdot\b \epsilon_{\lambda\b k}^* e^{-ik_z vt}}{k^2-\lambda \sigma_\chi k -(k_zv)^2-i\sigma k_z v}\,.\label{a23}
 \gal
 \end{subequations}
Substituting these equations along with \eq{a13} into \eq{a6} and \eq{a7} on derives \cite{Tuchin:2014iua}:
\bal
\b B(\b r, t)= &\int \frac{d^3k}{(2\pi)^3}\frac{qvke^{ik_z(z- vt)}e^{i\b k_\bot\cdot \b b}}{\left[ k^2-(k_zv)^2-i\sigma k_zv\right]^2-(\sigma_\chi k)^2}\nonumber\\
&\times \left\{ \left[ k^2-(k_zv)^2-i\sigma k_z v\right]\sum_\lambda\lambda\b\epsilon_{\lambda \b k}(\unit z\cdot \b\epsilon^*_{\lambda\b k}) + \sigma_\chi k\sum_\lambda\b\epsilon_{\lambda \b k}(\unit z\cdot \b\epsilon^*_{\lambda\b k})
\right\}\label{a24A}\,.
\gal
The explicit expressions for the polarization sums can be found in Appendix. 
In particular, the azimuthal component reads, upon using \eq{b16} and \eq{b18} and integrating over the angle between $\b k$ and $\b b$:
\bal
B_\phi(\b r, t)
= &\int \frac{dk_\bot k_\bot}{(2\pi)^2}\int_{-\infty}^\infty dk_z\frac{qvk_\bot e^{ik_z(z- vt)}J_1(k_\bot b) }{\left[ k^2-(k_zv)^2-i\sigma k_zv\right]^2-(\sigma_\chi k)^2} \left[ k^2-(k_zv)^2-i\sigma k_z v\right] \,.
\label{a24B}
\gal

Thus far the calculation has been completely general.  Now, employing the approximation $k\approx k_z$ the $k_\bot$-integral can be done by writing in \eq{a24B}
\bal
&\frac{k^2-(k_zv)^2-i\sigma k_z v }{\left[ k^2-(k_zv)^2-i\sigma k_zv\right]^2-(\sigma_\chi k_z)^2}=  
\frac{1}{2}\sum_{\lambda=\pm 1} \frac{1}{k_\bot^2+k_z^2/\gamma^2-i\sigma k_zv +\lambda\sigma_\chi k_z}\,.
\gal
The result is
\ball{a24C}
B_\phi(\b r, t)= &\frac{qv}{8\pi^2}\int_{-\infty}^{+\infty} dk_z e^{ik_z(z-vt)}\nonumber\\
&
\times \sum_{\lambda=\pm 1}
\sqrt{k_z^2/\gamma^2-i\sigma k_zv +\lambda\sigma_\chi k_z}\,K_1\!\!\left(b\sqrt{k_z^2/\gamma^2-i\sigma k_zv +\lambda\sigma_\chi k_z}\right)\,.
\gal
This equation is plotted in \fig{fig:2} for $t<\tau_c$. The other components of the magnetic and electric field can be computed in a similar way \cite{Tuchin:2014iua,Li:2016tel}.

Employing a stronger approximation $\sigma \ll k_\bot \ll k_z\ll \sigma\gamma$ and integrating in \eq{a24B} first over $k_z$ and then over $k_\bot$ obtains a simple formula \cite{Tuchin:2014iua,Li:2016tel}
\bal
B_\phi(\b r, t)
\approx  &\int \frac{dk_\bot k_\bot}{(2\pi)^2}\int_{-\infty}^{+\infty} dk_z\frac{qvk_\bot e^{ik_z(z- vt)}J_1(k_\bot b) }{\left[ k_\bot^2-i\sigma k_zv\right]^2-(\sigma_\chi k)^2} \left[ k_\bot^2-i\sigma k_z v\right] 
\nonumber\\
=& \frac{qb}{8\pi \zeta^2}\exp\left( -\frac{b^2\sigma}{4\zeta}\right)\left[ \sigma \cos\left( \frac{b^2\sigma_\chi}{4\zeta}\right)+\sigma_\chi \sin \left( \frac{b^2\sigma_\chi}{4\zeta}\right)\right]\,,\label{a24D}
\gal
 Eq.~\eq{a24D} is a very good approximation of \eq{a24C} in the relativistic heavy-ion collision kinematics.

\section{Electromagnetic field at  later times}\label{sec:b2}

To study the late time $t\gg \tau_c$ behavior of the electromagnetic field, one can regard the chiral conductivity $\sigma_\chi(t)$ as a random process and hence \eq{a18} becomes a stochastic equation describing time-evolution of the field amplitude with momentum $\b k$ and polarization $\lambda$. 
Introducing an auxiliary variable $x=  \Phi_{\lambda \b k}e^{\sigma t/2}$ one can cast \eq{a18} in the form
\ball{a26}
\ddot x(t)+\omega^2[1+\alpha \xi (t)]x(t)= \lambda k J_{\lambda \b k}(t)e^{\sigma t/2}\,,
\gal
where 
\ball{a27}
\omega^2= k^2-\frac{\sigma^2}{4}\,,\qquad
\alpha = -\frac{\lambda k}{\omega^2}\Sigma_\chi\,,\qquad \xi(t) = \frac{\sigma_\chi}{\Sigma_\chi}\,.
\gal
Eq.~\eq{a26} describes the one-dimensional harmonic oscillator with random frequency. It does not have an analytical solution. However, one can deduce from it a set of ordinary differential equations for the expectation value of the amplitude  moments \cite{VanKampen:1975}. In particular,  assuming $\alpha\ll 1$, the average value of $x$ satisfies the equation
\ball{a29}
\partial_t^2\aver{x(t)}+\frac{1}{2}c_2\alpha^2\omega \partial_t\aver{x(t)}+\left(1-\frac{1}{2}\alpha^2c_1\right)\omega^2\aver{x(t)}= \lambda k J_{\lambda \b k}(t)e^{\sigma t/2}(1+\alpha^2 c_0)\,.
\gal
where 
\begin{subequations}
\bal
&c_0(\omega)=\int_0^\infty  \mathcal{K}(\tau) \sin( \omega \tau)[1-\cos( \omega \tau)]d(\omega \tau)\,,\label{a32}\\
&c_1(\omega)=\int_0^\infty  \mathcal{K}(\tau)\sin(2 \omega \tau)d( \omega \tau)\,,\label{a33}\\
&c_2(\omega)=\int_0^\infty  \mathcal{K}(\tau)[1-\cos(2 \omega \tau)]d( \omega \tau)\,.\label{a34}
\gal
\end{subequations}
with the autocorrelation function $\mathcal{K}(\tau)= \aver{\xi(t)\xi(t-\tau)}$. Eq.~\eq{a29} can be converted into the equations for the average of the amplitude $\Phi_{\lambda \b k}$:
\ball{a40} 
\partial_t^2 \aver{\Phi_{\lambda \b k}}+\left(\sigma+\frac{\alpha^2}{2}c_2\omega\right) \partial_t\aver{ \Phi_{\lambda \b k}}+\left(\omega^2+\frac{\sigma^2}{4}-\frac{\alpha^2}{2}c_1\omega^2+\frac{\alpha^2}{4}c_2\sigma\omega\right)\aver{\Phi_{\lambda \b k}}\nonumber\\
= \lambda k J_{\lambda \b k}(t)(1+\alpha^2 c_0)\,.
\gal
The terms proportional to $\alpha^2$ represent contributions of the fluctuating chiral conductivity. Solution to \eq{a40} is
\ball{a41}
\aver{\Phi_{\lambda \b k}(t)}= \frac{qv \unit z \cdot\b \epsilon_{\lambda\b k}^*\lambda k(1+\alpha^2 c_0)e^{-ik_z vt}}{k^2 -(k_zv)^2-i\sigma k_z v+\alpha^2 Q(\omega) }\,.
\gal
 were a shorthand notation is used 
\ball{c2}
Q(\omega) =  \frac{1}{4}\left( c_2\sigma\omega-2ic_2k_zv\omega -2c_1\omega^2\right) \,.
\gal
 Substituting \eq{a41} into \eq{a6} yields  the magnetic field:
\ball{c1}
 \aver{\b B(\b r,t)}&= \sum_\lambda\int \frac{d^3k}{(2\pi)^3}e^{ik_z(z- vt)}e^{i\b k_\bot\cdot \b b}\frac{qv \b\epsilon_{\lambda \b k}(\unit z \cdot\b \epsilon_{\lambda\b k}^*)\lambda k(1+\alpha^2 c_0)}{k^2 -(k_zv)^2-i\sigma k_z v+\alpha^2 Q(\omega) }\,,
  \gal
The $z$-component of the magnetic field vanishes due to \eq{b11}. Its $b$ component vanishes when \eq{b19} is substituted into \eq{c1} and integrated over the azimuthal angle $\psi$.  Using \eq{b18} the azimuthal component of the magnetic field is 
\ball{c4}
 \aver{B_\phi(\b r,t)}&= \int \frac{d^3k}{(2\pi)^3}e^{ik_z(z- vt)}e^{i\b k_\bot\cdot \b b}\frac{(-i\b k\cdot \unit b) qv (1+\alpha^2 c_0)}{k^2 -(k_zv)^2-i\sigma k_z v+\alpha^2 Q(\omega) }\,.
  \gal

The electric field is obtained using \eq{a7}, \eq{a13} and \eq{a41}:
\ball{c5}
 \aver{\b E(\b r,t)}=&\int \frac{d^3k}{(2\pi)^3}e^{i\b k_\bot \cdot \b b-ik_z(vt-z)}
 \left\{ 
- \sum_\lambda\frac{\lambda}{k}
 \frac{(-ik_zv) qv \b\epsilon_{\lambda \b k}(\unit z \cdot\b \epsilon_{\lambda\b k}^*)\lambda k(1+\alpha^2 c_0)}{k^2 -(k_zv)^2-i\sigma k_z v+\alpha^2 Q(\omega) }
 -\frac{i\b k}{k^2\varepsilon_{k_zv}}
 \right\}
 \gal
In particular, employing \eq{b8} and \eq{b17} the non-vanishing components are 
\bal
 \aver{ E_z(\b r,t)}&=q\int \frac{d^3k}{(2\pi)^3}e^{i\b k_\bot \cdot \b b-ik_z(vt-z)}\frac{ik_z}{k^2}
 \frac{ k^2\left(v^2-\varepsilon_{k_zv}^{-1}\right)-\alpha^2 Q(\omega)\varepsilon_{k_zv}^{-1}+ \alpha^2 c_0k_\bot^2v^2}{k^2 -(k_zv)^2-i\sigma k_z v+\alpha^2Q(\omega)}\label{c7}\,,
\\
 \aver{ E_b(\b r,t)}&=-q\int \frac{d^3k}{(2\pi)^3}e^{i\b k_\bot \cdot \b b-ik_z(vt-z)}\frac{i\b k\cdot \unit b}{k^2}
 \frac{\left( k^2+\alpha^2 Q(\omega)\right) \varepsilon_{k_zv}^{-1}+ \alpha^2 c_0k_z^2v^2}{k^2 -(k_zv)^2-i\sigma k_z v+\alpha^2Q(\omega)}\,.\label{c8}
 \gal
 Thus, the direction of the \emph{average} electric and magnetic fields is the same as in the anomaly-free case. 

In the heavy-ion collision kinematics discussed in \sec{sec:k} $\omega\approx k_z$ which implies $\alpha \approx -\lambda\Sigma_\chi/k_z$ and  $\alpha^2 Q(\omega)\approx -\Sigma_\chi^2\left[i c_2(k_z)+c_1(k_z)\right]/2$. This allows taking the transverse momentum integrals in \eq{c4},\eq{c7} and \eq{c8} 
\begin{subequations}\label{D1}
 \bal
 \aver{B_\phi(\b r,t)}=& \frac{qv}{(2\pi)^2}\int_{-\infty}^{+\infty} dk_z(1+\alpha^2c_0)sK_1(bs) e^{-ik_z(vt-z)}\,,\label{d1}\\
\aver{E_z(\b r,t)}=& \frac{qi}{(2\pi)^2}\int_{-\infty}^{+\infty} dk_z k_z\left( v^2-\frac{1}{\e_{vk_z}}-\frac{\alpha^2}{4k_z^2\e_{vk_z}}Q(vk_z)\right) K_0(bs) e^{-ik_z(vt-z)}\,,\label{d2}\\
   \aver{E_b(\b r,t)}=& \frac{q}{(2\pi)^2}\int_{-\infty}^\infty dk_z\left(1+\frac{\alpha^2}{4k_z^2} Q(vk_z)\right)\frac{1}{\e_{vk_z}}sK_1(bs) e^{-ik_z(vt-z)}\,,\label{d3}
 \gal
where 
\ball{d5}
s^2= \frac{k_z^2}{\gamma^2}-ik_zv\sigma + \alpha^2 Q(vk_z)\,.
\gal
 \end{subequations}
Note that all three terms are kept in \eq{d5}, i.e.\ no assumption is made about the relationship between $k_z$ and $\sigma \gamma$ (i.e.\ condition \texttt{(iii)} is not imposed as no further analytical integration can be done anyway). As a result, \eq{d1} reduces to \eq{k10} when $\sigma= \Sigma_\chi=0$. Similarly \eq{d2},\eq{d3} also reduce to their corresponding classical free space expressions in this limit.

To estimate the numerical effect of the topological fluctuations on the electromagnetic field, consider the Ornstein-Uhlenbeck random process with the auto-correlation function 
$\mathcal{K}(\tau)=\exp(-\tau/\tau_c)$. The corresponding coefficients \eq{a32}--\eq{a34} read 
\ball{d7}
&c_0(\omega)= \frac{4(\tau_c\omega)^3}{1+4(\tau_c\omega)^2}\,,\qquad c_1(\omega)=\frac{2(\omega\tau_c)^2}{1+4(\omega\tau_c)^2}\,,\qquad 
c_2(\omega)= \frac{4(\tau_c\omega)^3}{1+4(\tau_c\omega)^2}\,.
\gal

\begin{figure}[ht]
\begin{tabular}{cc}
      \includegraphics[height=4.5cm]{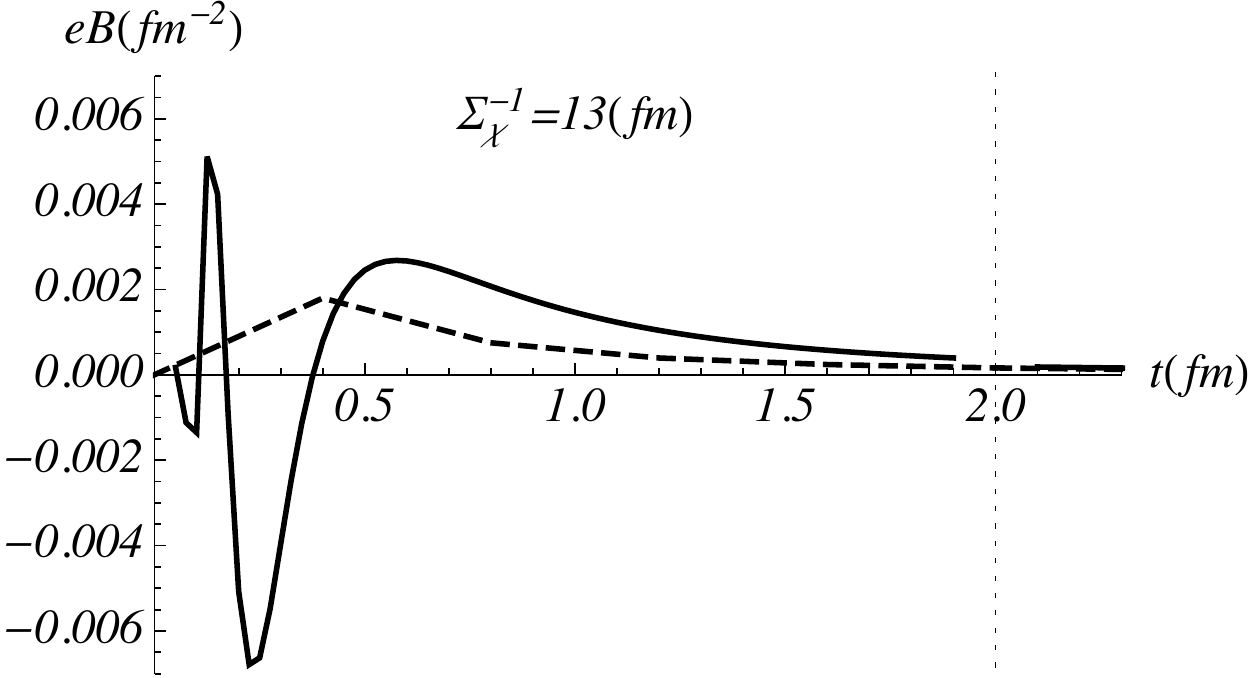} &
      \includegraphics[height=4.5cm]{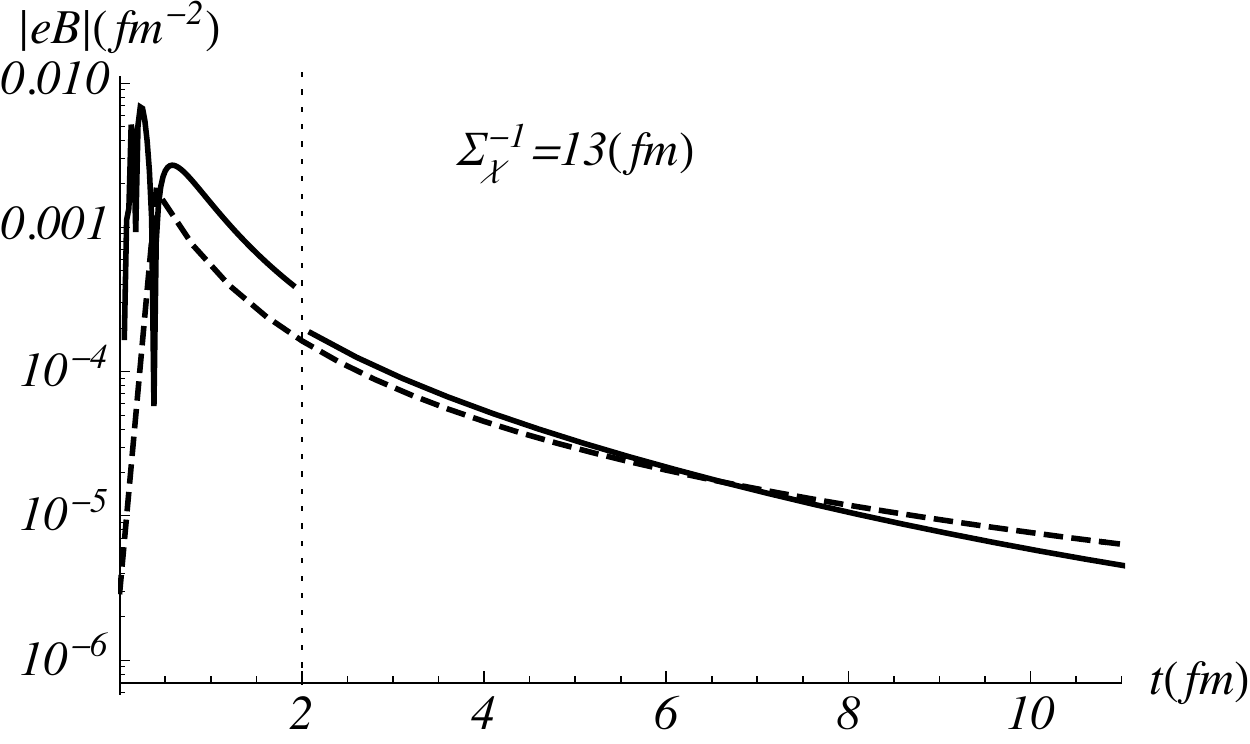}\\
            \includegraphics[height=4.5cm]{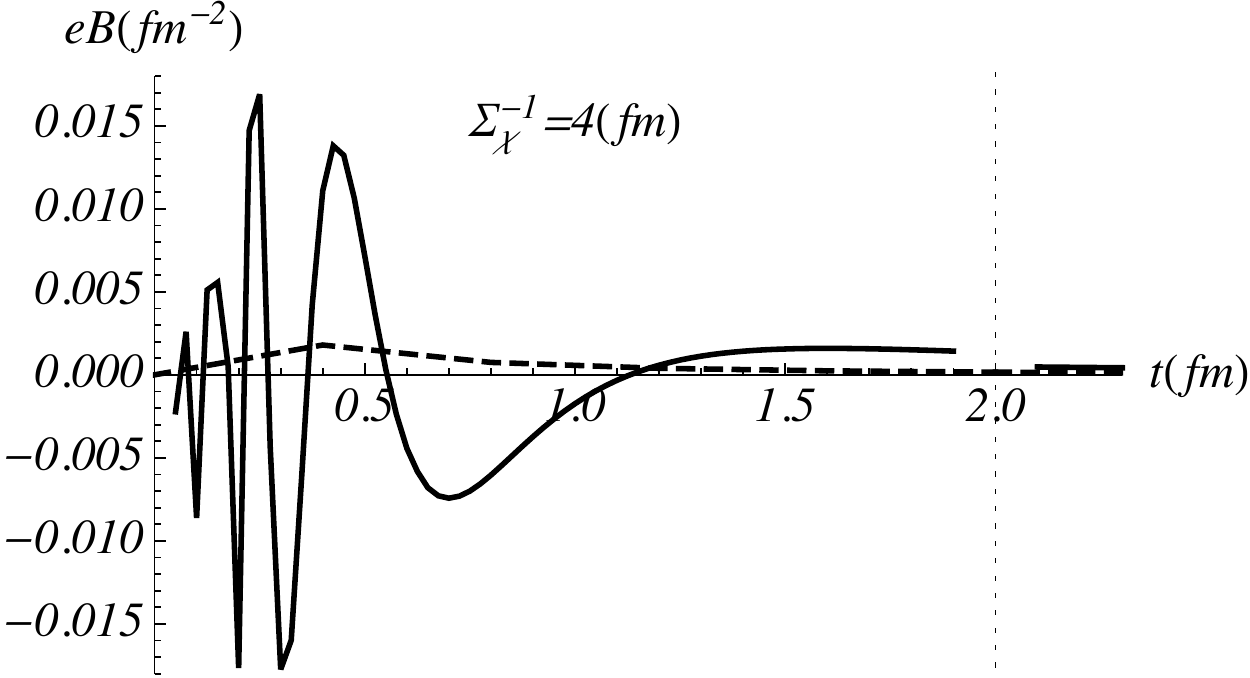} &
      \includegraphics[height=4.5cm]{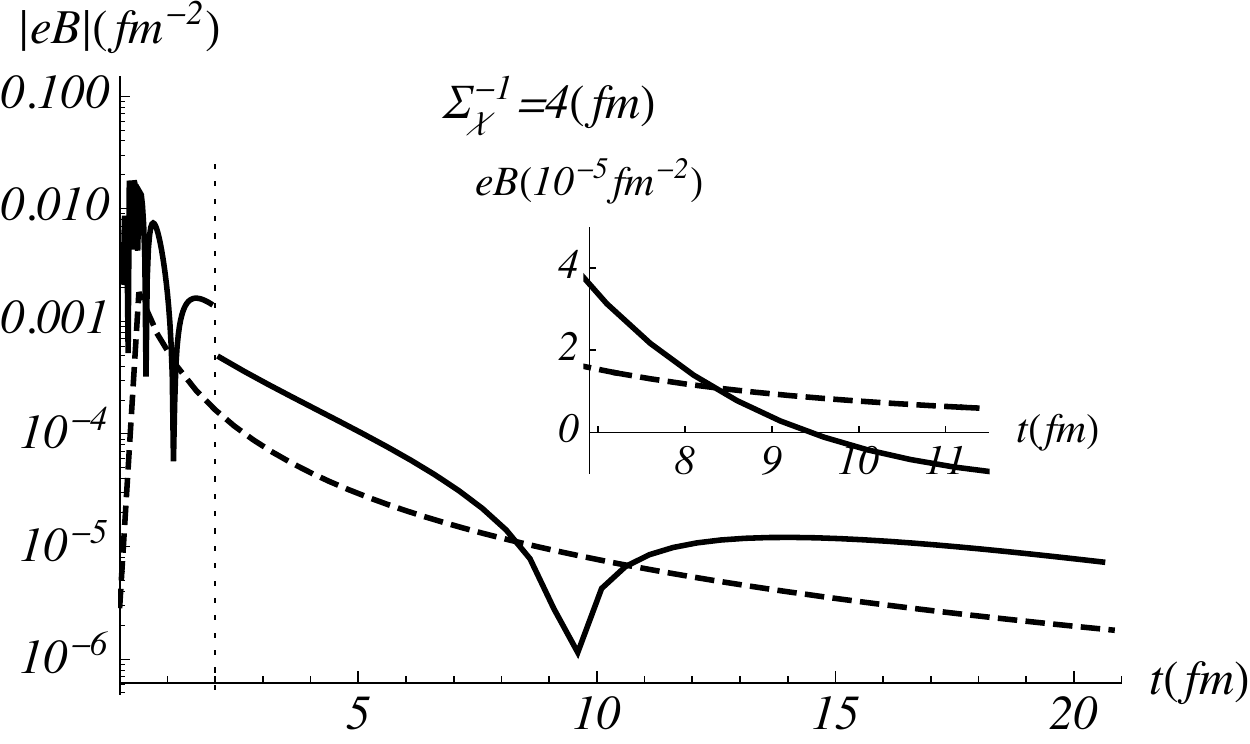}
      \end{tabular}
  \caption{Azimuthal component of magnetic field (left panels) and its absolute value (right panels) at $\tau_c=2$~fm (vertical dotted line).  Solid lines:  $\aver{\sigma^2_\chi(t)}^{1/2}=\Sigma_\chi\neq 0$ as indicated on each panel, dashed lines: $\aver{\sigma^2_\chi(t)}=0$. The minima of the right panels are the zeros of the field at which it reverses its direction, as shown in the inset in the lower right panel.  Other parameters: $\gamma=100$, $\sigma^{-1}=34$~fm, $b=7.4$~fm, $z=0$. }
\label{fig:2}
\end{figure}

Magnetic field \eq{a24C} and \eq{d1} is plotted in \fig{fig:2} for different values of $\Sigma_\chi$ and $\tau_c=2$~fm. It must be stressed that neither of these quantities is presently constrained by experiment. Their values in \fig{fig:2} are chosen for the presentation clarity. 
The field discontinuity at $t=\tau_c$ is mere reflection of the fact that neither solution can be trusted at $t=\tau_c$. 
The main feature is that the magnetic field oscillates when $\Sigma_\chi$ is sufficiently large compared to $\sigma$.

\section{Summary}\label{sec:s}

In this paper the electromagnetic field of a fast electric charge in chiral medium is computed in two cases: in \sec{sec:b} when the chiral conductivity is constant  and in \sec{sec:b2} when it is random. In the former case the previous result derived in \cite{Tuchin:2014iua} is reproduced, while in the later one a new result given by \eq{c1},\eq{c7} and \eq{c8} is derived. In the relativistic heavy-ion collisions kinematics the field expressions reduce to \eq{D1}. \fig{fig:2} is the graphic representation of the magnetic field produced in a typical heavy-ion collision by a single valence quark. One observes that the field oscillations at early times, first observed in \cite{Tuchin:2014iua}, may persist at later times at large chiral conductivity. If the chiral conductivity is indeed that large, observation of the chiral magnetic \cite{Kharzeev:2007jp,Fukushima:2008xe} and associated effects \cite{Kharzeev:2015znc} in relativistic heavy-ion collisions becomes especially challenging.

Throughout the paper the topological charge density has been assumed to be spatially homogeneous \cite{Zhitnitsky:2014ria,Kharzeev:2007tn}. In practice this might not be a good approximation if more than one CP-odd domain is produced in a single heavy-ion collision. The impact of spatial and temporal variations as well  as the quantum interference effects deserve a dedicated analysis. Furthermore, throughout the paper it is assumed that the electrical conductivity is constant. This is clearly not the case in a realistic heavy-ion collisions. One should be mindful of these limitations when considering the phenomenological applications of the results of this work.

\acknowledgments
This work  was supported in part by the U.S. Department of Energy under Grant No.\ DE-FG02-87ER40371.

\appendix
\section{Polarization sums}\label{appA}

Let $\unit \xi$, $\unit \eta$ and $\unit k$ form a right-handed basis. The polarization vectors are given by
\ball{b0}
\b\epsilon_{\lambda\b k}= \frac{\unit \xi+i\lambda \unit \eta}{\sqrt{2}}\,.
\gal
And let  $k_z$ $k_\bot$ and $\psi$  be cylindrical coordinates of vector $\b k$ in the $\unit x$, $\unit y$, $\unit z$ basis i.e.\ 
\bal
\b k &= k_z\unit z+k_\bot \cos\psi\unit x+ k_\bot \sin \psi \unit y\,, \label{b1}\\
\unit \eta & = \frac{1}{k}(-k_\bot \unit z+k_z\cos\psi \unit x+k_z\sin\psi \unit y)\,, \label{b2}\\
\unit \xi &= \sin\psi \unit x-\cos\psi \unit y\,. \label{b3}
\gal
Then 
\bal
&\b\epsilon_{\lambda\b k}\cdot \unit z= -\frac{i\lambda k_\bot}{\sqrt{2}k}\,, \label{b5}\\
& \b\epsilon_{\lambda\b k}\cdot \unit x=\frac{1}{\sqrt{2}}(\sin\psi+i\lambda k_z\cos\psi)\,,\label{b6}\\
&  \b\epsilon_{\lambda\b k}\cdot \unit y=\frac{1}{\sqrt{2}}(-\cos\psi+i\lambda k_z\sin\psi)\,.\label{b7}
\gal
It can be easily shown that 
\bal
&\sum_\lambda |\b\epsilon_{\lambda\b k}\cdot \unit z|^2 =\frac{k_\bot^2}{k^2}\,,\label{b8}\\
&\sum_\lambda\b\epsilon_{\lambda\b k}\cdot \unit x\, \b\epsilon_{\lambda\b k}^*\cdot \unit z =-\frac{k_zk_\bot}{k^2}\cos\psi\,,\label{b9}\\
&\sum_\lambda \b\epsilon_{\lambda\b k}\cdot \unit y\, \b\epsilon_{\lambda\b k}^*\cdot \unit z =-\frac{k_zk_\bot}{k^2}\sin\psi\,,\label{b10}\\
&\sum_\lambda |\b\epsilon_{\lambda\b k}\cdot \unit z|^2\lambda =0\,,\label{b11}\\
&\sum_\lambda\b\epsilon_{\lambda\b k}\cdot \unit x\, \b\epsilon_{\lambda\b k}^*\cdot \unit z\lambda =\frac{ik_\bot}{k}\sin\psi\,,\label{b12}\\
& \sum_\lambda\b\epsilon_{\lambda\b k}\cdot \unit y\, \b\epsilon_{\lambda\b k}^*\cdot \unit z\lambda =-\frac{ik_\bot}{k}\cos\psi\,. \label{b13}
\gal
These formulas are particular cases of the more general expressions for the polarization sums: 
\bal
&\sum_\lambda (\epsilon^i_{\lambda\b k})^*  \epsilon^j_{\lambda\b k}= \delta^{ij}-\frac{k^ik^j}{k^2}\label{b14}\,,\\
& \sum  _\lambda \lambda(\epsilon^i_{\lambda\b k})^*  \epsilon^j_{\lambda\b k} =\frac{i \varepsilon^{ij\ell} k_\ell}{k}\,.\label{b15}
\gal

Define the unit vectors of the cylindrical coordinates with respect to $\unit z$-axis as
\bal
&\unit \phi = -\sin\phi\unit x +\cos\phi \unit y\,,\\
& \unit b = \cos\phi \unit x+\sin\phi \unit y \,,
\gal
where $\phi$ is the angle between the radial vector $\b b$ in the $xy$-plane and the $x$-axis. Then
\bal
&\sum_\lambda\b\epsilon_{\lambda\b k}\cdot \unit \phi\, \b\epsilon_{\lambda\b k}^*\cdot \unit z =-\frac{k_zk_\bot}{k^2}\sin(\psi-\phi)\,,\label{b16}\\
&\sum_\lambda\b\epsilon_{\lambda\b k}\cdot \unit b\, \b\epsilon_{\lambda\b k}^*\cdot \unit z =-\frac{k_zk_\bot}{k^2}\cos(\psi-\phi)\,,\label{b17}\\
&\sum_\lambda\lambda \b\epsilon_{\lambda\b k}\cdot \unit \phi\, \b\epsilon_{\lambda\b k}^*\cdot \unit z =-\frac{ik_\bot}{k}\cos(\psi-\phi)\,,\label{b18}\\
&\sum_\lambda\lambda \b\epsilon_{\lambda\b k}\cdot \unit b\, \b\epsilon_{\lambda\b k}^*\cdot \unit z =\frac{ik_\bot}{k}\sin(\psi-\phi)\,. \label{b19}
\gal




\begin{thebibliography}{80}
  

\bibitem{Tuchin:2014iua} 
  K.~Tuchin,
  ``Electromagnetic field and the chiral magnetic effect in the quark-gluon plasma,''
  Phys.\ Rev.\ C {\bf 91}, no. 6, 064902 (2015)
  
  
\bibitem{Li:2016tel} 
  H.~Li, X.~l.~Sheng and Q.~Wang,
  ``Electromagnetic fields with electric and chiral magnetic conductivities in heavy ion collisions,''
  Phys.\ Rev.\ C {\bf 94}, no. 4, 044903 (2016)
  
\bibitem{Arnold:1997gh} 
  P.~B.~Arnold and L.~G.~Yaffe,
  ``Effective theories for real time correlations in hot plasmas,''
  Phys.\ Rev.\ D {\bf 57}, 1178 (1998)
  
\bibitem{Zhitnitsky:2014ria} 
  A.~R.~Zhitnitsky,
  ``The topological long range order in QCD. Applications to heavy ion collisions and cosmology,''
  EPJ Web Conf.\  {\bf 95}, 03041 (2015)
  
  
\bibitem{Kharzeev:2007tn} 
  D.~Kharzeev and A.~Zhitnitsky,
  ``Charge separation induced by P-odd bubbles in QCD matter,''
  Nucl.\ Phys.\ A {\bf 797}, 67 (2007)
  

  
\bibitem{Tuchin:2019gkg} 
  K.~Tuchin,
 ``Time-evolution of magnetic field in hot nuclear matter with fluctuating topological charge,''
  arXiv:1911.01357 [hep-ph].
  
\bibitem{Wilczek:1987mv} 
  F.~Wilczek,
  ``Two Applications of Axion Electrodynamics,''
  Phys.\ Rev.\ Lett.\  {\bf 58}, 1799 (1987).

\bibitem{Carroll:1989vb} 
  S.~M.~Carroll, G.~B.~Field and R.~Jackiw,
  ``Limits on a Lorentz and Parity Violating Modification of Electrodynamics,''
  Phys.\ Rev.\ D {\bf 41}, 1231 (1990).

\bibitem{Sikivie:1984yz} 
  P.~Sikivie,
  ``On the Interaction of Magnetic Monopoles With Axionic Domain Walls,''
  Phys.\ Lett.\ B {\bf 137}, 353 (1984).
 
  
  

 
\bibitem{Aarts:2007wj}
  G.~Aarts, C.~Allton, J.~Foley, S.~Hands and S.~Kim,
  ``Spectral functions at small energies and the electrical conductivity in
  hot, quenched lattice QCD,''
  Phys.\ Rev.\ Lett.\  {\bf 99}, 022002 (2007)

\bibitem{Ding:2010ga} 
  H.-T.~Ding, A.~Francis, O.~Kaczmarek, F.~Karsch, E.~Laermann and W.~Soeldner,
  ``Thermal dilepton rate and electrical conductivity: An analysis of vector current correlation functions in quenched lattice QCD,''
  Phys.\ Rev.\ D {\bf 83}, 034504 (2011)
   
\bibitem{Amato:2013oja} 
  A.~Amato, G.~Aarts, C.~Allton, P.~Giudice, S.~Hands and J.~I.~Skullerud,
  ``Transport coefficients of the QGP,''
  PoS LATTICE {\bf 2013}, 176 (2014)
  
  
\bibitem{VanKampen:1975} 
N.G.~Van~Kampen, "Stochastic Differential Equations", Phys. Rept. {\bf 24}, 173, (1976).


  
\bibitem{Kharzeev:2007jp} 
  D.~E.~Kharzeev, L.~D.~McLerran and H.~J.~Warringa,
  ``The Effects of topological charge change in heavy ion collisions: 'Event by event P and CP violation',''
  Nucl.\ Phys.\ A {\bf 803}, 227 (2008)

\bibitem{Fukushima:2008xe} 
  K.~Fukushima, D.~E.~Kharzeev and H.~J.~Warringa,
  ``The Chiral Magnetic Effect,''
  Phys.\ Rev.\ D {\bf 78}, 074033 (2008)


\bibitem{Kharzeev:2015znc} 
  D.~E.~Kharzeev, J.~Liao, S.~A.~Voloshin and G.~Wang,
  ``Chiral magnetic and vortical effects in high-energy nuclear collisions—A status report,''
  Prog.\ Part.\ Nucl.\ Phys.\  {\bf 88}, 1 (2016)
 

  

  
  
  

\end{thebibliography}
\end{document}